# VLBI DATA PROCESSING ON CORONAL RADIO-SOUNDING EXPERIMENTS OF MARS EXPRESS

Maoli Ma,[1,2] Guifré Molera Calvés,[3] Giuseppe Cimò,[4] Peijin Zhang,[5] Xiong Ming,[6] Peijia Li,[1]
Pradyumna Kummamuru,[3] zhanghu Chu,[1] Tianyu Jiang,[1] Bo Xia,[1] Kondo Tetsuro,[1] Fengxian Tong,[1]
Pablo de Vicente,[7] Jonathan Quick,[8] Hua Zhang,[9] and Zhong Chen[1,10]

[1]Shanghai Astronomical Observatory, 80 Nandan Road, Shanghai,China

[2]Shanghai Astronomical Observatory, Key Laboratory of Radio Astronomy, Chinese Academy of Sciences, 10 Yuanhua Road, Nanjing,
JiangSu 210033, China

[3]University of Tasmania, Private Bag 37, Hobart TAS 7001, Australia

[4]Joint Institute for VLBI ERIC, Oude Hoogeveensedijk 4, 7991 PD Dwingeloo, The Netherlands

[5]University of Science and Technology of China, 96 Jinzhai Road, Hefei, China

[6]National Space Science Center, NO.1 Nanertiao, Zhongguancun, Haidian district, Beijing

[7]Centro de Desarrollos Tecnológicos, Observatorio de Yebes, Guadalajara, Spain

[8]Hartebeesthoek Radio Astronomy Observatory, PO Box 443, Krugersdorp, South Africa

[9]Xinjiang Astronomical Observatory, CAS 150 Science 1-Street, Urumqi, China

[10]National Basic Science Data Center, Building No.2, 4, Zhongguancun South 4th Street, Haidian District, Beijing 100190, China

## ABSTRACT

The ESA's Mars Express solar corona experiments were performed at two solar conjunctions in the years 2015 and 2017 by a number of radio telescopes in the European VLBI Network. This paper presents the methods to measure the frequency and phase fluctuations of the spacecraft radio signal, and the applications to study the characteristics of the plasma turbulence effects on the signal at a single station and at multiple stations via cross-correlation. The power spectra of the frequency fluctuations observed between 4.9 and 76.3 $R_s$ have a power-law shape close to a Kolmogorov spectrum over the frequency interval $\nu_{lo} < \nu < \nu_{up}$, where the nominal value of $\nu_{lo}$ is set to 3 mHz and $\nu_{up}$ is in the range of $0.03 \sim 0.15$ Hz. The RMS of the frequency fluctuations is presented as a function of the heliocentric distance. Furthermore, we analyse the variations of the electron column density fluctuations at solar offsets 4.9 $R_s$ and 9.9 $R_s$ and the cross-correlation products between the VLBI stations. The power density of the differential fluctuations between different stations decreases at $\nu < 0.01$ Hz. Finally, the fast flow speeds of solar wind $> 700$ km s$^{-1}$ are derived from the cross-correlation of frequency fluctuations at $\nu < 0.01$ Hz. The fast flow speeds of solar wind correspond to the high heliolatitude of the coronal region that the radio rays passed. The VLBI observations and analysis methods can be used to study the electron column density fluctuations and the turbulence at multiple spatial points in the inner solar wind by providing multiple lines of sight between the Earth and the spacecraft.

## 1. INTRODUCTION

The Very Long Baseline Interferometry (VLBI) observation of a spacecraft inside the solar system has been applied to a broad range of research interests in recent years. The combination of VLBI measurables and Doppler spacecraft tracking have been demonstrated in a number of applications in planetary and space science, including tracking of the VEGA balloons for determining the wind field in the atmosphere of Venus (Ekonomov et al. 1986), the wind measurements during the VLBI tracking of the descent and landing of the Huygens Probe in the atmosphere of Titan (M.K.Bird et al. 2005), the characterization of the interplanetary plasma with Venus Express data (Molera Calvés et al. 2014), the detection of interplanetary coronal mass ejection (Molera Calvés et al. 2017), and the characterization of planetary atmosphere and/or ionosphere (Bocanegra-Bahamón et al. 2019). The work presented here provides the characterization of density turbulence in the extended solar corona by means of observations of spacecraft with VLBI antennas during solar conjunction.

Strong turbulence is generated in the solar wind and affects the propagation of radio signals. The turbulence plays an essential role in several aspects of plasma behavior. While considerable progress has been made, the production of energy injection, the solar wind acceleration and the heating of the extended solar wind remain as significant unsolved problems (Chen 2016). One way to address this issue is via observations of deep space probes during solar conjunction. The spacecraft carrier signal experiences refraction, absorption and scattering as it passes through the visible limb of the Sun. The measurements of frequency and phase fluctuations of the carrier signal are related to the change in the electron column density along the ray path. A number of articles have reported the electron column density fluctuations and the turbulence characteristics using the radio signal transmitted by interplanetary spacecraft. Chashei et al. (2005) and Efimov et al. (2008) reported unusual power spectra with a sharp break or flatter spectra inside 20 solar radii ($R_s$) with NASA's (National Aeronautics and Space Administration) Galileo spacecraft, and the NASA and the ESA's (European Space Agency) Ulysses spacecraft. Pätzold et al. (2012) identified the dense coronal streamers and several coronal mass ejections by observing the radio transmission of ESA's Mars Express (MEX), Venus Express, and Rosetta, and determined the total electron content as a function of heliocentric distance. Miyamoto et al. (2014) explored the radial variations of the amplitude and the energy flux of compressive waves in the solar corona using JAXA's (Japan Aerospace Exploration Agency) Akatsuki spacecraft at heliocentric distances of $1.5 \sim 20.5$ $R_s$. Wexler et al. (2019a) created a stacked, magnetically structured slab model that incorporated both propagating slow density waves and advected spatial density variations to explain the observed frequency fluctuations from Messenger 2009 and Helios $1975 \sim 1976$ superior conjunctions. A further study of the coronal energy transport was presented combining global magnetohydrodynamic (MHD) models with radio sounding observations with Akatsuki spacecraft at heliocentric distances of $1.6 \sim 1.86$ $R_s$ (Wexler et al. 2019b).

Compared with the intensity scintillation measured for natural sources, which can provide information in the order of the Fresnel region (Tokumaru et al. 2010, 2011), the observations of spacecraft during solar conjunction by separated stations provide a tool to study the inhomogeneity and the convective velocity of the irregularities at the large scale of turbulence (Woo 1977; Richard & Paul 1992). Yakovlev et al. (1989) reported the cross-correlation of frequency and phase fluctuations from NASA Venera-15 and -16 spacecraft behind the Sun recorded by two stations with a separation of 7000 km. They found the most abrupt increase in solar wind speed occurs in the range $10 \sim 25$ $R_s$. N.A.Armand et al. (2003) developed the theory of frequency fluctuations in turbulent coronal plasma. The convective velocity of irregularities from $7 \sim 30$ $R_s$ were obtained through simultaneous observations of Galileo by NASA's Deep Space Network stations Goldstone, Canberra and Madrid. Hence, the wide coverage of VLBI antennas around the world has the potential to study the correlation of the scintillation and solar wind velocity profile in the spatial scale of the distance between the telescopes, the so-called baseline length.

In this paper, we further combine Doppler and VLBI techniques to present different data processing and analysis methods applied to radio observations of the MEX spacecraft at a solar conjunction. The advantage of our methods is that the frequency and phase fluctuations of the carrier signal along different line of sights (LOSs), as well as the fluctuations among different baselines, could be obtained simultaneously. The fluctuations are then used to analyse the density turbulence at each single station and their cross-correlation at multi-stations as well. The observational setup is introduced in Sect. 2. A short description of the frequency and phase measurements is introduced in Sect. 3. The fluctuation analysis for the single stations is presented in Sect. 4. The cross-correlation of the scintillation at the multi-stations, including the variation of the electron column density, and the estimation of the convective velocity of the solar wind irregularities are presented in Sect. 5. Finally, our conclusions are presented in Sect. 6.

## 2. EXPERIMENTAL SETUP

The radio observations of the solar corona were carried out using ground-based VLBI telescopes of the European VLBI Network, we measured the radio signal of MEX during two superior conjunctions in 2015 and 2017. The first solar conjunction took place during the descending phase of the solar cycle, where solar radio flux at 10.7 cm ($F_{10.7}$) were in the range of $130 \sim 100$ s.f.u. (solar flux units)[1], and the second conjunction occurred near the minimum solar activity, with $F_{10.7}$ varying in the range of $70 \sim 80$ s.f.u.

The observations are summarized in Table 1. We carried out 17 observations of MEX in 2015 with the Sheshan 25-metre telescope located in Shanghai (abbreviated as Sh). The solar offset distance was in the range of $11.2 \sim 76.3$ $R_s$. Six observations were carried out in 2017. Two of them were recorded simultaneously at five different EVN stations, i.e., Hartebeesthoek (two antennas), Yebes, Noto, Urumqi and Tianma (abbreviated as Hh/Ht, Ys, Nt, Ur and T6, respectively) on July 29 and August 3 in 2017 when solar offsets were 4.9 $R_s$ and 9.9 $R_s$.

**Table 1**. Specifications of the observations: Date, time, solar offset, stations, number of scans, ingress or egress. Each scan was 19-minute or 15-minute

| Date | Time | Solar offset $R_s$ | Stations | No. of scans | Ingress/Egress |
|------|------|--------------------|----------|--------------|----------------|
| 2015-03-30 | 05:20-09:19 | 72.4 | Sh | 12 | I |
| 2015-04-02 | 02:45-06:38 | 69.9 | Sh | 11 | I |
| 2015-04-10 | 08:20-09:39 | 62.9 | Sh | 4 | I |
| 2015-04-17 | 00:20-01:39 | 56.9 | Sh | 3 | I |
| 2015-04-20 | 05:00-07:19 | 54.1 | Sh | 6 | I |
| 2015-05-04 | 06:40-08:19 | 41.2 | Sh | 4 | I |
| 2015-05-08 | 07:40-09:19 | 37.4 | Sh | 5 | I |
| 2015-05-18 | 04:00-06:19 | 27.8 | Sh | 11 | I |
| 2015-05-21 | 01:40-02:39 | 25.0 | Sh | 3 | I |
| 2015-05-21 | 04:20-05:59 | 24.9 | Sh | 5 | I |
| 2015-06-25 | 01:04-03:19 | 11.2 | Sh | 5 | E |
| 2015-06-26 | 00:00-02:19 | 12.3 | Sh | 15 | E |
| 2015-07-02 | 07:20-08:59 | 18.8 | Sh | 3 | E |
| 2015-07-09 | 01:40-03:59 | 26.2 | Sh | 9 | E |
| 2015-08-08 | 00:40-02:59 | 59.4 | Sh | 11 | E |
| 2015-08-14 | 03:20-05:39 | 66.2 | Sh | 8 | E |
| 2015-08-23 | 00:20-01:19 | 76.3 | Sh | 3 | E |
| 2015-09-06 | 05:40-07:19 | 92.2 | Sh | 4 | E |
| 2017-06-30 | 02:00-02:39 | 31.1 | Sh | 2 | I |
| 2017-07-06 | 08:45-09:59 | 24 | Sh | 4 | I |
| 2017-07-07 | 07:40-08:19 | 23.4 | Sh | 2 | I |
| 2017-07-21 | 06:20-08:39 | 8.1 | Sh | 7 | I |
| 2017-07-29 | 06:00-13:10 | 4.9 | Ht,Ys,Nt,Ur,T6 | 8 | E |
| 2017-08-03 | 06:00-13:00 | 9.9 | Hh,Ys,Nt,Ur,T6 | 11 | E |

We observed the signal transmitted in X-band (8.4 GHz). The setup of the observations was a three-way mode radio link. The transponder on board was working in closed-loop with one of the antennas of the ESA tracking stations, New Norcia or Cebreros, while VLBI antennas observed the downlink signals. The stability of the signal in closed-loop mode allows precise measurements of the signal frequency and phase (Molera Calvés et al. 2014). The down-converted radio signals were recorded using VLBI data acquisition systems: Mark5B and VDIF, with 4 and 16 MHz bandwidth, and two-bit encoding. To make recording and re-pointing of the antenna simple, each observation was split into 19-minute or 15-minute scans.

## 3. THE FREQUENCY AND PHASE MEASUREMENTS

Under strong interplanetary scintillation, the S/C signal recorded by ground stations mainly includes the Doppler shift, the fluctuations due to the solar plasma, the radio noise radiated from the sun, antenna noise, etc. In order to extract the fluctuations caused by solar wind plasma, one needs to eliminate the Doppler shift caused by the relative motion between the spacecraft and the ground station.

In this work, the compensation of the Doppler shift is achieved by two different methods. The first method follows the pipeline processing developed by the Planetary Radio Interferometry and Doppler Experiments (PRIDE) group at the Joint Institute for VLBI ERIC (JIVE) (Molera Calvés et al. 2014). We firstly estimate topocentric frequency $f_c$ of the main carrier through instantaneous Doppler measurement (Zheng et al. 2013). These detections allow us to derive the Doppler variation of the spacecraft signal through the entire scan. The Doppler shift is modelled using an (n-1)-th order frequency polynomial fit, which is equivalent to an n-th order phase-stopping polynomial fit. We use a 6-th order phase polynomial fit to agree with the data processing in Molera Calvés et al. (2014). After the phase stop, the carrier signal is converted into near-stationary signals. We use the phase lock loop (PLL) to track the residual frequency and phase, $\Delta f_c$ and $\Delta \varphi_c$. Then a low-order fit is applied on the residuals to obtain the fluctuations, $\delta f_c$ and $\delta \varphi_c$ for scintillation analysis. The sampling time of the output products is 1 s. The loop bandwidth of PLL is set to 20 Hz for all the sessions. Subtracting the Doppler shift through a polynomial fit on the measured topocentric frequency to obtain the fluctuations is a useful and efficient method. However, the power spectra density analysis on the scintillation shows that it will introduce some uncertainty at the low-frequency spectra, which is extremely important to study the large scale turbulence (Wexler et al. 2020; Jensen et al. 2018). In order to analyse the scintillation on VLBI multi-station observations, we have developed the second method to eliminate the Doppler shift using detailed ephemerides modeling. Ma et al. (2017, 2020) called this technique *local* correlation.

The schematic representation of the observation of a spacecraft by two ground stations during a radio sounding experiment is illustrated in Fig. 1(a). The relative positions of Sun (S), Earth (E), and spacecraft (S/C) form the sounding plane S-E-S/C. The solar offset of the radio ray path R is the distance from the center of the Sun to the solar proximate point P along the line-of-sight between Earth and spacecraft. Corresponding solar proximate points along the ray paths from the spacecraft to the two widely-separated ground stations A and B are denoted $P_1$ and $P_2$. $P_1P_2$ is the actual spatial separation of the radio ray paths from the spacecraft to two ground stations A and B. The projection of $P_1P_2$ onto the solar radial is $\Delta R$,

$$\Delta R = SP_1 \cos(\alpha) - SP_2 \cos(\beta), \tag{1}$$

where $\alpha$ and $\beta$ are the angle between $SP_1$ and $SP$, $SP_2$ and $SP$, respectively. $\tau_A$ and $\tau_B$ in Fig. 1(a) are the radio propagation delay from the spacecraft to the two spaced stations. In the three-way radio link, $\tau_A$ and $\tau_B$ include both the uplink and downlink range trips. During solar conjunction observations, the plasma crosses different LOSs and causes the different signal scintillation. The baseline orientated north-south is sensitive to the solar wind at high heliolatitudes, while east-west to the low heliolatitudes (Rao et al. 2008).

Figure 1(b) presents the algorithm of the local correlation. $s_A(t)$ and $s_B(t)$ are the received signals at the stations A and B. The method compensates the Doppler shift with a signal propagation model constructed by the delay model $\tau_{A,B}$ and the uplink transmitted frequency $f_t$ of the carrier. $\tau_{A,B}$ are derived from the spacecraft's *a-priori* ephemerides (Ma et al. 2020; Huang et al. 2014). After the Doppler compensation, the carrier signal is converted into near-stationary signals with a slow change in phase. We then use PLL and a low order fit to obtain the frequency and phase fluctuations, $\delta f_c^{A,B}$ and $\delta \varphi_c^{A,B}$. The differential fluctuations at the baseline are obtained as the difference of fluctuations between two separated stations, thus $\delta f^{AB}(t) = \delta f_c^A(t) - \delta f_c^B(t)$, $\delta\varphi^{AB}(t) = \delta\varphi_c^A(t) - \delta\varphi_c^B(t)$. The local correlation method differs from the classical VLBI correlator which provides the phase and frequency fluctuations at each single station, as well as the differential fluctuations in the spatial scale of the baseline, while the VLBI correlator for spacecraft can provide the cross-correlation phase along a baseline.

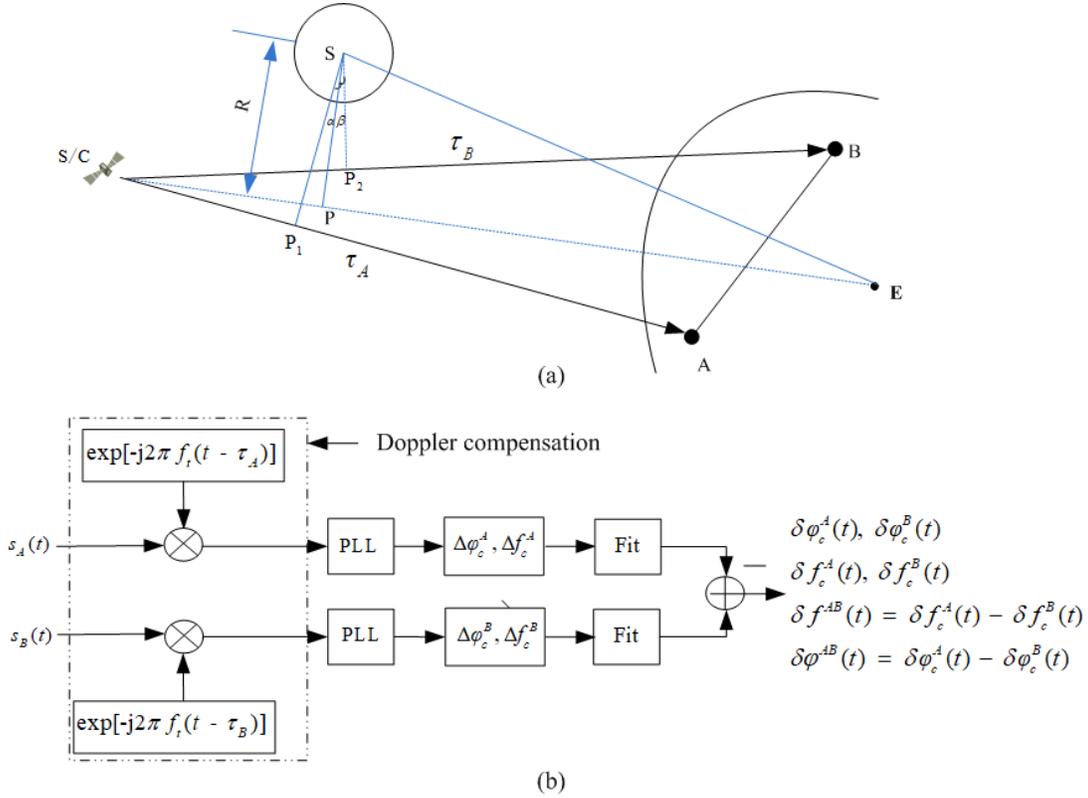

**Figure 1.** (a) Schematic representation of the observation of the spacecraft by two spaced stations during a radio sounding experiment. (b) The algorithms of local correlation. $s_A(t)$ and $s_B(t)$ are the received signals at the two different stations. The local correlation compensates for the Doppler shift of spacecraft using the spacecraft's priori ephemerides.

In the three-way Doppler, the transmitted frequency from the uplink station is needed to reconstruct the model signal. As we did not have the uplink transmitted frequency, the PRIDE-method is used in the single-dish observations in 2015 and 2017. The local correlation method is used in the VLBI observations on July 29 and August 3, 2017.

## 4. FLUCTUATION ANALYSIS AT SINGLE STATION

In this section, we present the characteristics of the solar wind turbulence with observational data from a single station. Data utilized are presented in Table 1.

### 4.1. Spectral analysis of the frequency fluctuations

We use the power spectra to determine the radial dependence of the intensity of the frequency fluctuations. The spectra of the frequency fluctuations are obtained using the average of all individual spectra calculated with a standard FFT-1024 algorithm on each individual scan (19 min or 15 min). Fig. 2 shows the averaged power density spectra, $G_f(\nu)$, of the frequency fluctuations at several different solar offset. The spectra have a near power-law shape close to the Kolmogorov-spectrum (Kolmogorov 1991) in a frequency range $[\nu_{lo}, \nu_{up}]$ that corresponds to the inertial range. $\nu_{lo}$ is firstly determined by the comparison of the fluctuations, $\delta f_c$ and $\delta\varphi_c$ from the two different Doppler compensation methods (Appendix A). The power density spectra of $\delta f_c$ and $\delta\varphi_c$ from the two methods are exactly the same when $\nu > 3$ mHz. A small difference appears below 3 mHz. The difference comes from the complex dynamical motion of spacecraft that cannot be fitted by a polynomial model. In the single-dish observations when the local correlation is not available, the $\nu_{lo}$ is determined by the comparison of the spectra of $\delta f_c$ using different polynomial fit orders for Doppler compensation (Appendix B). The power density spectra of the $\delta f_c$ are also different when $\nu < 3$ mHz. Therefore, though the low-frequency plasma fluctuations are sensitive to the polynomial fit, we can ensure the validity when $\nu > \nu_{lo}$, $\nu_{lo} \sim 3$ mHz, which is consistent with the value used in Molera Calvés et al. (2014).

$\nu_{up}$ depends on the intensity of the turbulence and the noise. In Fig. 2, besides the instrument noise, the radio frequency signals emitted from the sun are also an important noise source[2], and enhance with the decline of solar offset distance as well. When using the same filter bandwidth in measurement, the energy competition between turbulence and noise bring in the variation of $\nu_{up}$, which generally decreases with the decline in the intensity of the turbulence. We estimate $\nu_{up}$ through a visual inspection of the spectrum to find out the turning point between the turbulence and noise. Here, $\nu_{up}$ is in the range of $0.03 \sim 0.15$ Hz.

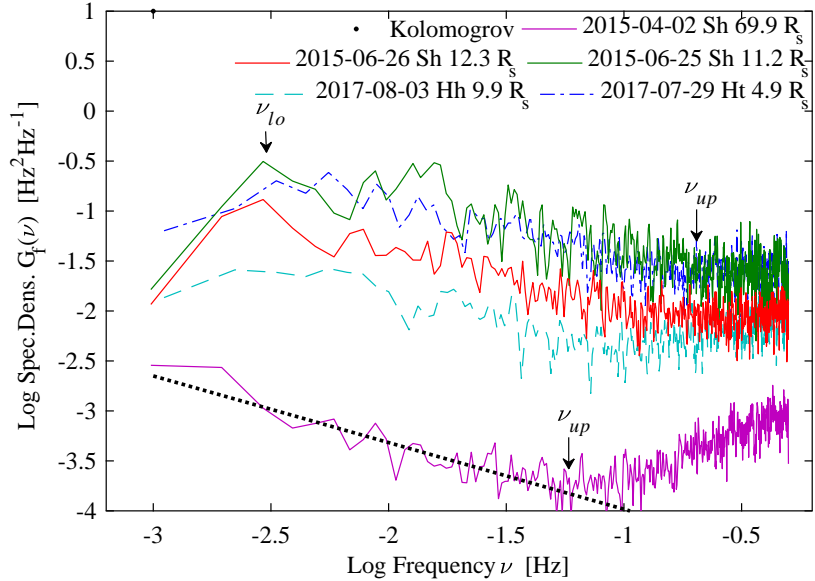

**Figure 2.** Power spectra of frequency fluctuation at different solar offsets. The spectra at 4.9 $R_s \sim 69.9$ $R_s$ have a near power-law shape close to that of the Kolmogorov-spectrum in a frequency range $[\nu_{lo}, \nu_{up}]$. The noise dominates the spectra when $\nu > \nu_{up}$. Data from July 29 and August 3, 2017, were obtained by the local correlation. The PRIDE approach was used on all the rest.

According to N.A.Armand et al. (2003) and Efimov et al. (2010), the spectrum has a pronounced maximum and the corresponding turnover frequency can be used to determine the outer scale of turbulence. The turnover frequency is very low, 0.1 mHz in Efimov et al. (2002) and Wohlmuth et al. (2001). This means that we need a continuous observation with at least $\sim 10000$ seconds. In this work, the outer scale of turbulence is constrained by the relatively short length of the time series segments (1000 seconds). Therefore, longer observations are necessary to determine the outer scale of the plasma inhomogeneities in the future.

### 4.2. Radial dependence of the Doppler scintillation parameters

We characterize the temporal spectra of frequency fluctuations using the RMS intensity of the frequency fluctuations $\sigma_f$ and the spectral index $\alpha_f$. $\sigma_f^2$ is obtained by numerical integration of the power spectral density within the inertial range, $[\nu_{lo}, \nu_{up}]$ (Wexler et al. 2020).

$$\sigma_f^2 = \int_{\nu_{lo}}^{\nu_{up}} G_f(\nu) d\nu \qquad [\text{Hz}^2] \qquad (2)$$

The spectral index $\alpha_f$ is determined by fitting the spectra data within the inertial range. The nominal value of $\nu_{lo}$ in this study is 3 mHz (see, Appendix). The overall radial dependence of the $\sigma_f$ is shown in Fig. 3. The $\sigma_f$ tends to increase with the decreasing heliocentric distance. This is consistent with previous observations (Imamura et al. 2005; Efimov et al. 2008).

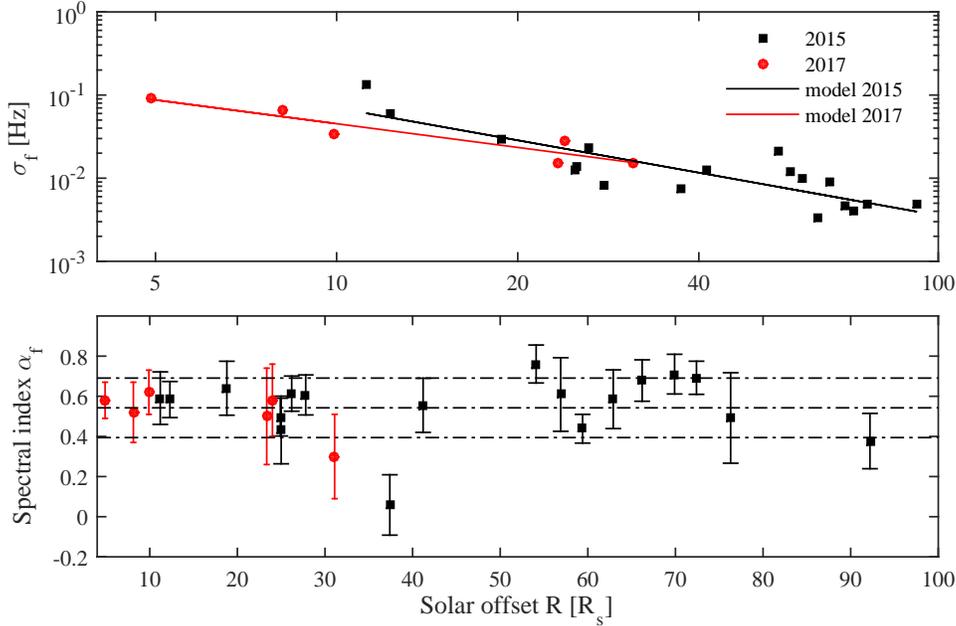

**Figure 3.** The RMS intensity of the frequency fluctuations and spectral index versus solar offset distance. The circles represent the data from 2015 and the dots represent the data from 2017.

The radial dependence of $\sigma_f$ is approximated by a following power-law function:

$$\sigma_f = B \cdot (R/R_s)^{-b}, \qquad (3)$$

which corresponds to a linear function at the double logarithmic scale. The following values are derived from the fit of $\sigma_f$ with $R$ in 2015 and 2017 respectively:

$$B = 1.37 \text{ Hz}, b = 1.29 \qquad (2015)$$
$$B = 0.40 \text{ Hz}, b = 0.95 \qquad (2017) \qquad (4)$$

The radial falloff exponents $b$ are 1.29 in 2015 and 0.95 in 2017. Efimov et al. (2013) reported the results from different spacecraft (including MEX) from the years 2004, 2006 and 2008. According to their results, the X-band power-law indices were ranging $1.50 \sim 1.93$ when the solar offsets were between 2.4 and 46.74 $R_s$. We obtained lower indices compared with Efimov et al. (2013). One possible reason is that the minimum solar offset is 11.2 $R_s$ in 2015, and we lack the samples at a smaller solar offset. In 2017, the RMS intensity of the frequency fluctuations $\sigma_f$ at solar offset 4.9 $R_s$ are lower than the value at 11.2 $R_s$ in 2015. A comparison of the frequency fluctuations with the Large Angle and Spectrometric Coronagraph (LASCO) images is presented in the next subsection. The mean of the measured spectral index is 0.58. The classic 3-D spatial turbulence spectral index p is 11/3 (Kolmogorov 1991), and the theoretical spectral index for frequency fluctuations is $\alpha_f = $ p - 3 = 2/3 (Yakovlev 2017). Our measurement is close to the theoretical value expected from a Kolmogorov-type turbulence spectrum (Kolmogorov 1991; Yakovlev 2017).

### 4.3. Comparison with LASCO images

The strongest scintillation usually occurs at the closest approach of the ray path to the Sun. However, as shown in Fig. 2 and 3, the frequency fluctuations at 11.2 $R_s$ in 2015 are stronger than the fluctuations at 4.9 $R_s$ and 9.9 $R_s$ in 2017. The measurements suggest that stronger sources of perturbation were present on the data sets at 11.2 R$_s$ in 2015.

We compared the fluctuations on June 25/26, 2015 and July 29 and August 3, 2017, with the large-scale structure of the solar corona observed in LASCO. LASCO, an instrument onboard the NASA SOHO mission, has two coronagraphs that provide good coverage of the solar corona from 1.5 to 32 R$_s$. The C3 coronagraph makes images of the solar corona from about 3.5 to 32 R$_s$, and the C2 coronagraph from about 1.5 to 6 R$_s$ (Brueckner et al. 1995). Fig. 4 shows the coronal structures on June 25/26, 2015 at 01:54 UT, and the corona on July 29 and August 3, 2017, around 10:00 UT. The position of Mars is indicated by a white arrow in each image. On June 25/26, 2015, *a number of bright coronal rays emanated radially from the coronal hole and crossed the LOS*. The LOS was fully engulfed in the bright streamer structure. On July 29 and August 3 in 2017, the position angles of MEX measured from the solar north were 342 and 311 degrees[3]. The corresponding heliolatitudes of the coronal region passing by the radio rays were 70º and 40º, respectively. The LOSs were closer to the northern coronal regions, where the rays were much fainter. Therefore, the stronger frequency fluctuations at 11.2 R$_s$ in 2015 are consistent with the LASCO images.

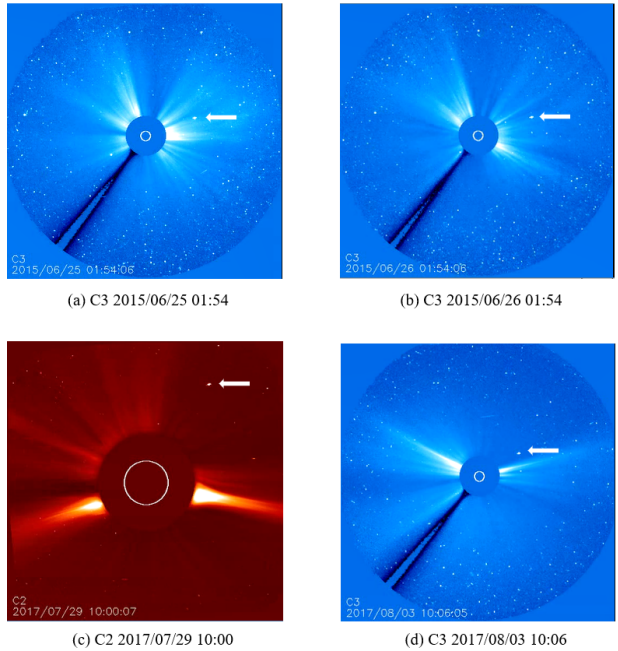

(a) C3 2015/06/25 01:54

(b) C3 2015/06/26 01:54

(c) C2 2017/07/29 10:00

(d) C3 2017/08/03 10:06

**Figure 4.** LASCO images on June 25/26, 2015 and July 29, August 3, 2017. Mars is marked with a white arrow. On June 25/26, 2015, a number of bright coronal rays emanated radially from the coronal hole and crossed the LOSs. On July 29 and August 3 in 2017, the rays were fainter.

## 5. CROSS-CORRELATION OF THE FLUCTUATIONS

The EVN stations observed the MEX on July 29 and August 3 in 2017, when the solar offset distances were 4.9 R$_s$ and 9.9 R$_s$. As shown in Table 1, Hh/Ht, Ys, Nt, Ur and T6 participated in the observations. Hh/Ht is located in South Africa, with the latitude ∼ 25º S, the other four stations are located between 31º ∼ 44º N. T6 is located in east Asia with the longitude ∼ 121º E. The combination of the five stations forms the baselines orientated at different north-south and east-west directions, with the longest baseline up to ∼ 10000 km, which will be sensitive to solar winds at both high or low heliolatitudes. We first use the instantaneous Doppler measurement method (Ma et al. 2020; Zheng et al. 2013) to measure the topocentric frequency and the corresponding formal error, then the formal error is used to evaluate the validity of the frequency detection. As shown in Fig. 5, the performances of formal error at different stations are quite different. Hh/Ht and T6 are relatively stable during the entire sessions. However, due to the limitation at visibility, T6 participated in only 5 minutes in the observations on July 29 and 75 minutes on August 3. Other three stations have many abnormal points. The effect of the sun in the sidelobes is under investigation. Ur data could not be used for the analysis, and only part of Ys data on July 29 could be decoded.

[3] https://ssd.jpl.nasa.gov/horizons.cgi

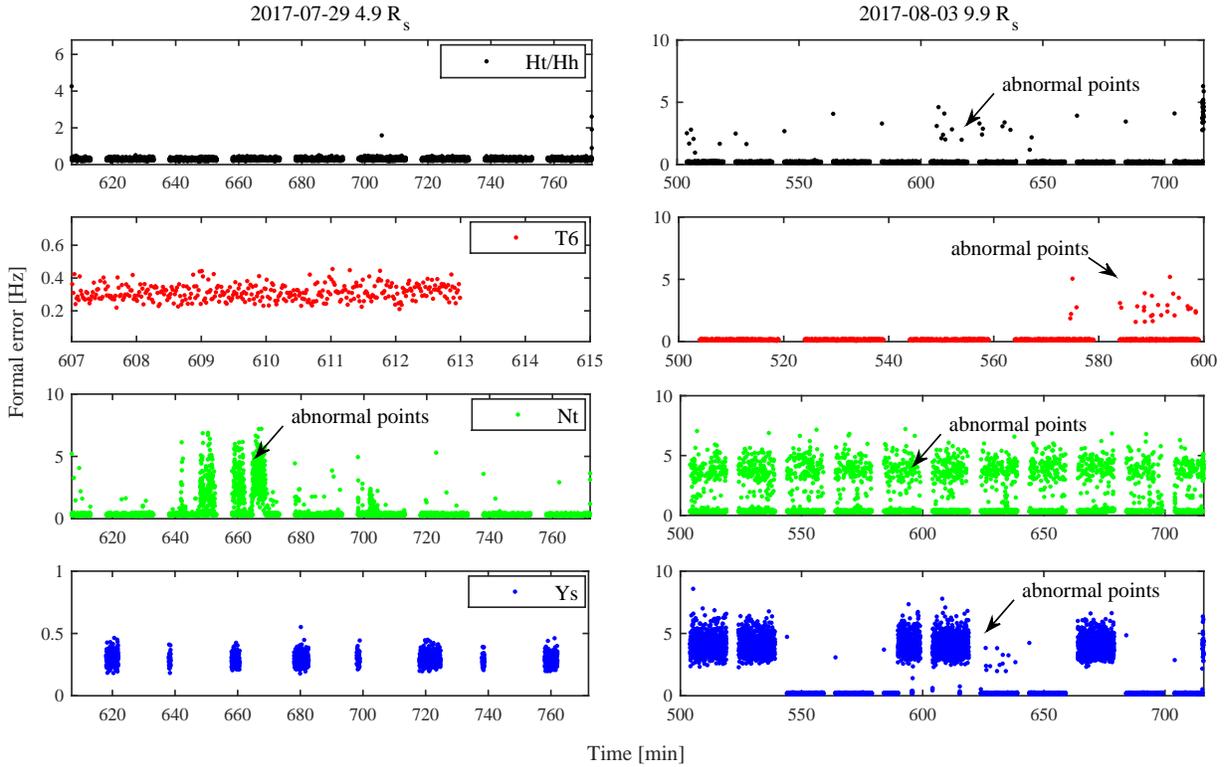

**Figure 5.** Frequency formal error. Left: 4.9 R_s. Right: 9.9 R_s. Hh/Ht and T6 are relatively stable during the whole sessions. Limited by the visibility, T6 only participated in a 5-mins observation on July 29 and a 75-mins observation on August 3, 2017. There are many abnormal breaks in the data. The receiving system's response to the sun in the sidelobes is under investigation.

After the preliminary detection of the signal, we used the local correlation to measure the frequency and phase fluctuations. A significant phase jumps occurs at many abnormal points (Fig. 5). We carefully delete these phase jumps through a cautious comparison of the residuals of frequency $\Delta f_c(t)$ and phase $\Delta \varphi_c(t)$ among different stations. A few inapparent phase jumps can be smoothed by the correlation. Fig. 6 shows the $\Delta \varphi_c(t)$ of the multi-station after we removed the phase jumps. On July 29 we have 9 valid scans for HT, 8 for Nt, 1 for T6, and incomplete 8 scans for Ys. On August 3, we extract 11 scans for Hh, 5 for T6, and 5 scans for Ys.

### 5.1. *Electron column density fluctuations at ground stations*

We reduce the contributions of the residual phase $\Delta \varphi_c(t)$ due to the error in the trajectory by a second-order polynomial fit to the phase time series and subtracted it from the original time series to obtain the phase fluctuation $\delta \varphi_c$. The time-dependent change in electron column density with an undetermined offset, $\Delta N_e$, has a relation with $\delta \varphi_c$ as (Ando et al. 2015)

$$\Delta N_e(t) = \frac{c f_c \delta \varphi_c(t)}{2\pi \alpha},$$ (5)

where $c$ is the speed of light, $\alpha = e^2/8\pi^2\varepsilon_0 m_e \approx 40.3 m^3 s^{-2}$ and $e$, $\varepsilon_0$, and $m_e$ are the elementary charge, dielectric constant in vacuum, and electron mass, respectively. Hence, the differential electron column density fluctuations between two stations, A and B, is given by

$$\Delta N_e^{AB}(t) = \Delta N_e^A(t) - \Delta N_e^B(t) = \frac{c f_c \delta \varphi^{AB}(t)}{2\pi \alpha},$$ (6)

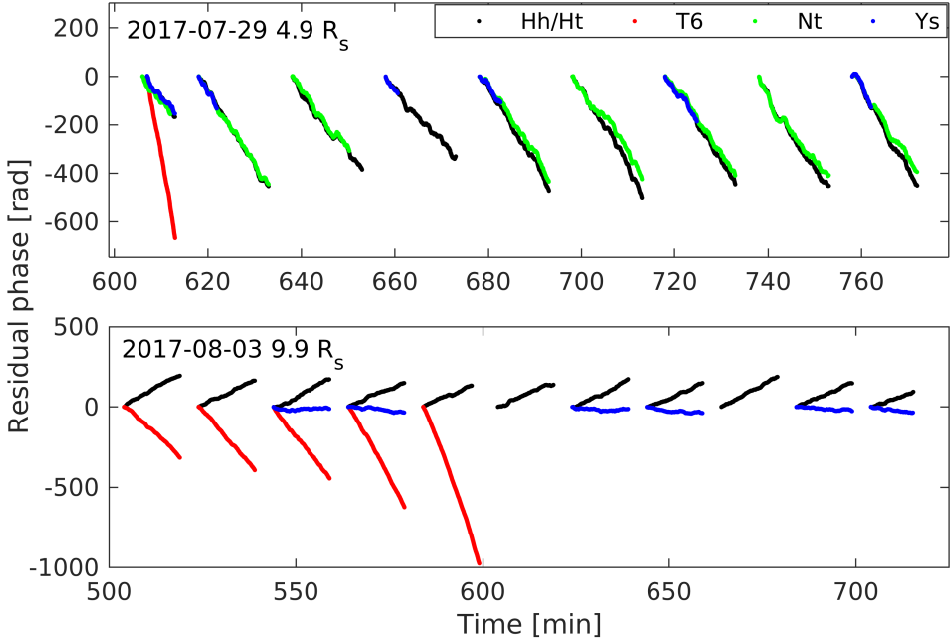

**Figure 6.** Residual phase relative to the prior orbit of multi-station on July 29 and August 3 in 2017. The significant phase jumps from the abnormal points were deleted.

where $\Delta N_e^A(t)$ and $\Delta N_e^B(t)$ are the density fluctuations at two seperate stations A and B, $\delta\varphi^{AB}(t)$ is the differential phase fluctuations between stations A and B. The examples of the $\Delta N_e$ from each station and their differential fluctuations $\Delta N_e^{AB}$ time series for multi-station with solar offset of 4.9 $R_s$ and 9.9 $R_s$ are presented in Fig. 7. The variations of $\Delta N_e$ at each station are similar. This means that the electron column density fluctuations along spatially separated ray paths are correlated. The correlation of $\Delta N_e^A$ and $\Delta N_e^B$ is in accordance with the Taylor hypothesis (Taylor 1938). The error for $\Delta N_e$ is mainly caused by the abnormal points in Fig. 5. Although they are smoothed by the local correlation, they affect the fluctuations to a certain extent (see Sect. 5.2). Random errors in $\Delta N_e$ also arise from fluctuations of the Earth's ionosphere and atmosphere, and the thermal noise in the ground receiver. Bocanegra-Bahamón et al. (2019) evaluated the uncertainties of the frequency residual from thermal noise, antenna mechanical noise, propagation noise, and the noise budget of the X-band Doppler detections of the VLBI stations. They were within the range of $3.7 \sim 11.8$ mHz. Compared with the strong fluctuations at 4.9 $R_s$ and 9.9 $R_s$ ($\sigma_f \approx 0.1$Hz in Fig. 3), these errors can be ignored. The standard deviation (STD) of $\Delta N_e$ and $\Delta N_e^{AB}$ are summarized in Table 2. For the purpose of comparing data at different times and from different stations, scans with a duration of less than 15 mins were omitted. The STD of $\Delta N_e^{AB}$ decreases significantly due to the correlation of $\Delta N_e^A$ with $\Delta N_e^B$.

A power density spectrum analysis is applied to the electron column density fluctuations to detect the cross-correlation of the fluctuations (Fig. 7 bottom). The spatial power spectra of $\Delta N_e$ at a single station follows the Kolmogorov scaling. Comparing with $\Delta N_e$, the power density of $\Delta N_e^{AB}$ decrease significantly when $\nu < 0.01$ Hz, which strongly suggests that the fluctuations at different stations are correlated in this range. The corresponding spatial scale is larger than $v_c/0.01$ (km) where $v_c$ is the solar wind velocity (Efimov et al. 2013). Therefore, the scale of the correlation of the phase fluctuations is commensurable with the large scale of turbulence. $\Delta N_e^{AB}$ at different baselines also reflect the inhomogeneity of the electron density in the spatial projection direction. Compared with the STD of $\Delta N_e^{AB}$ from different baselines in Table 2, the values from Hh-T6 and T6-Ys are larger than those from Hh-Ys. In Fig. 7, the spectral density of $\Delta N_e^{AB}$ from Hh-T6 and T6-Ys is also larger than the values from Hh-Ys. This means that the correlation along east-west baselines are weaker than north-south baselines. This is consistent with the heliolatitudes of the region crossing the LOSs.

### 5.2. *Estimation of the convective velocity of the solar wind irregularities*

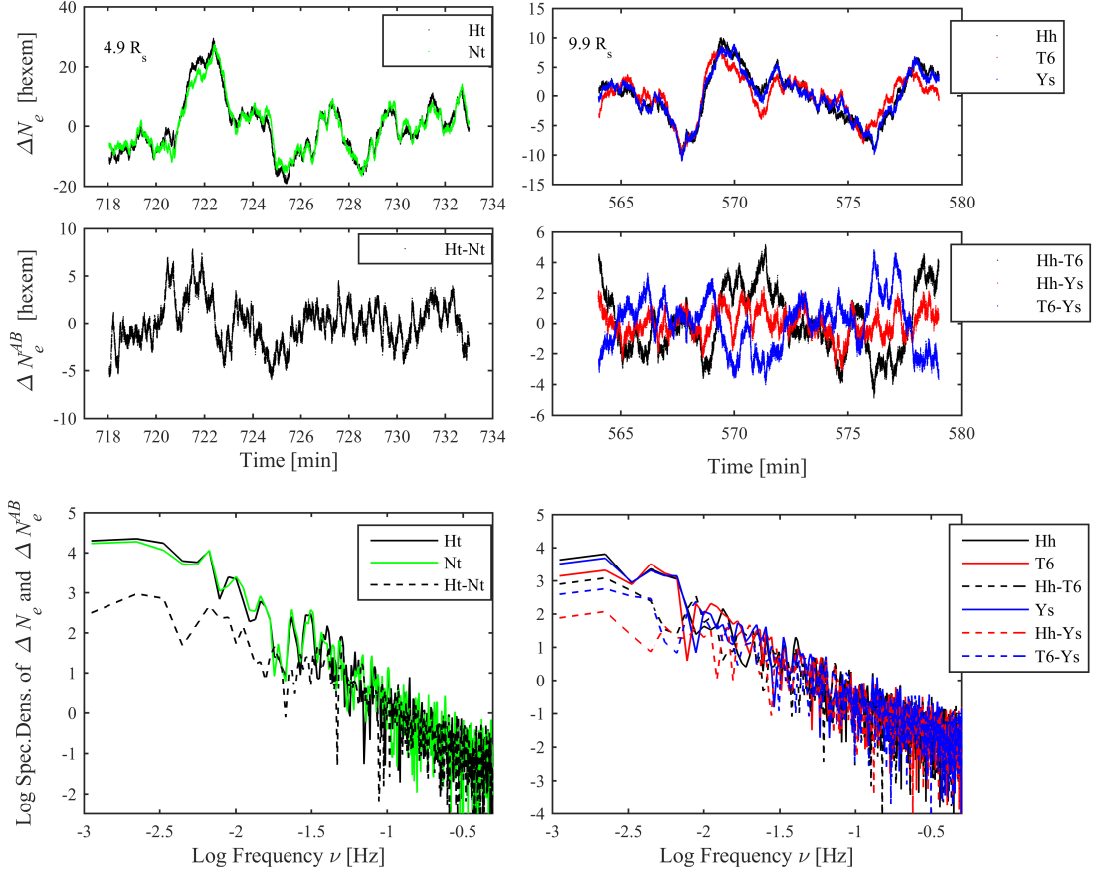

**Figure 7.** Upper panel: Electron column density fluctuations $\Delta N_e$ from each single station at a solar offset of 4.9 R$_s$ on July 29, 2017 and 9.9 R$_s$ on August 3, 2017. Middle panel: The corresponding differential electron column density $\Delta N_e^{AB}$ in the spatial scale of the baseline length. Lower panel: Power density spectra of $\Delta N_e$ and $\Delta N_e^{AB}$. Comparing with $\Delta N_e$, the power density of $\Delta N_e^{AB}$ decrease significantly when $\nu < 0.01$ Hz.

Based on the hypothesis of frozen-in plasma turbulence, N.A.Armand et al. (2003) proposed a method to estimate the convective velocity of solar wind irregularities by cross-correlating the frequency fluctuations from two different stations. The time lag obtained by the cross correlation is assumed to be the time taken by the solar wind to move the distance equal to the radial projection of the actual spatial separation of the radio ray paths between the two ground stations and spacecraft. Hence, the velocity of the solar wind irregularities, $v_c$, is given by

$$v_c = \frac{\Delta R}{\Delta \tau},\tag{7}$$

where $\Delta \tau$ is the time lag, and $\Delta R$ is calculated using Eq.(1). Experimental studies show that this method to estimate the solar wind velocity yields satisfactory results if the maximum correlation amplitude exceeds 0.6 and the radial projection $\Delta R$ is several thousand kilometers (Yakovlev 2014; Efimov et al. 2010). The scale of correlation is extremely important to study the cross-correlation of the frequency fluctuations among different LOSs. Fig. 7 shows that the correlation happened at $\nu < 0.01$ Hz. However, the few outliers in each scan (Fig. 5), which are smoothed by local correlation, will affect the correlation of the fluctuations on a large scale. One example is the first scan on July 29, 2017, at baseline Ht-Nt in Table 2. The power density of the differential phase and frequency do not decrease at $\nu < 0.01$ Hz (Fig. 8). We eliminate such types of scans through a cautious inspection of the power density spectra. The scans with a duration of less than 15 minutes are dropped, as well as the data sets with $\Delta R < 1000$ km, such as the data at the baseline of Hh–T6 on August 3, where $\Delta R$ varies from -1000 km to + 1000 km. The cutoff frequency $\nu_c$ is set to 0.01 Hz and 0.02 Hz, respectively.

**Table 2.** Standard deviation of the electron column density $\Delta N_e$ and the differential electron column density $\Delta N_e^{AB}$: Date, time, solar offset and baseline, $\Delta N_e^A$ is the electron column density from the first station at the baseline, $\Delta N_e^B$ is the electron column density from the second station at the baseline, and $\Delta N_e^{AB}$ is the differential electron column density.

| Date | Time [a] | Solar offset | Baseline | STD of $\Delta N_e^A$ | STD of $\Delta N_e^B$ | STD of $\Delta N_e^{AB}$ |
|---|---|---|---|---|---|---|
| | [min] | [$R_s$] | | [hexem] | [hexem] | [hexem] |
| 2017-07-29 | 625 | 4.9 | Ht - Nt | 7.10 | 7.55 | 3.61 |
| 2017-07-29 | 685 | 4.9 | Ht - Nt | 9.04 | 9.31 | 3.11 |
| 2017-07-29 | 725 | 4.9 | Ht - Nt | 9.67 | 8.97 | 2.30 |
| 2017-07-29 | 745 | 4.9 | Ht - Nt | 10.77 | 12.57 | 4.39 |
| 2017-08-03 | 511 | 9.9 | Hh - T6 | 3.12 | 3.59 | 1.35 |
| 2017-08-03 | 531 | 9.9 | Hh - T6 | 4.08 | 3.55 | 1.90 |
| 2017-08-03 | 551 | 9.9 | Hh - T6 | 4.56 | 4.06 | 2.00 |
| 2017-08-03 | 571 | 9.9 | Hh - T6 | 4.34 | 3.59 | 2.05 |
| 2017-08-03 | 591 | 9.9 | Hh - T6 | 4.56 | 4.08 | 3.16 |
| 2017-08-03 | 551 | 9.9 | Hh - Ys | 4.58 | 4.07 | 1.25 |
| 2017-08-03 | 571 | 9.9 | Hh - Ys | 4.34 | 4.08 | 0.89 |
| 2017-08-03 | 631 | 9.9 | Hh - Ys | 4.12 | 3.81 | 1.13 |
| 2017-08-03 | 651 | 9.9 | Hh - Ys | 3.50 | 3.45 | 0.88 |
| 2017-08-03 | 691 | 9.9 | Hh - Ys | 2.95 | 3.17 | 0.92 |
| 2017-08-03 | 551 | 9.9 | T6 - Ys | 4.06 | 4.06 | 1.33 |
| 2017-08-03 | 571 | 9.9 | T6 - Ys | 3.08 | 4.08 | 1.67 |

[a] The average time at each scan.

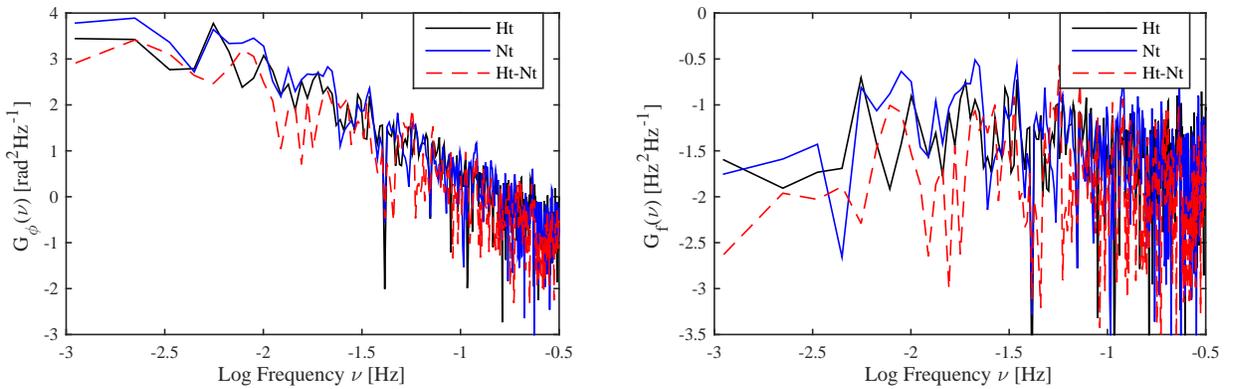

**Figure 8.** The power density spectra of phase (left) and frequency (right) fluctuations at 625 min, 2017-07-29, Ht-Nt. The power density of differential phase and frequency does not decrease at $\nu < 0.01$ Hz due to some abnormal outliers, which indicates the fluctuations does not correlate at $\nu < 0.01$ Hz.

The results of cross-correlation are presented in Fig. 9 and Table 3. Fig. 9 shows the correlation function of the frequency fluctuations from different scans and baselines on July 29 and August 3. The fluctuations $\delta f_c(t)$ are obtained from the 1st order fit on the residual frequency $\Delta f_c(t)$. The time lags are -4 s and -3 s for two individual scans at the baseline Ht–Nt on July 29, -6 s, -3 s and -3 s for other three individual scans at the baseline Hh–Ys and T6–Ys on August 3. The correlation coefficients are usually above 0.95, and a little lower (0.92) for T6–Ys. We also analyse the influence of the fitting order on the cross-correlation. The lags don't change on July 29 when using different the 2nd and 3rd order fits, and vary $\pm 1$ s between different orders on August 3. We finally take the mean time-lag among the different fit orders. The error on the lags is set to $\pm$ 0.5 s, corresponding to the time resolution. The variation of the radial projection $\Delta R$ is within 200 km in 15 minutes. Therefore we take the average value of $\Delta R$ in each scan. The corresponding convective velocity of solar wind irregularities are 950$\pm$120 km s$^{-1}$ and 1233$\pm$220 km s$^{-1}$ on July 29, 735$\pm$70 km s$^{-1}$, 1080$\pm$150 km s$^{-1}$ and 1000$\pm$140 km s$^{-1}$ on August 3 at different periods.

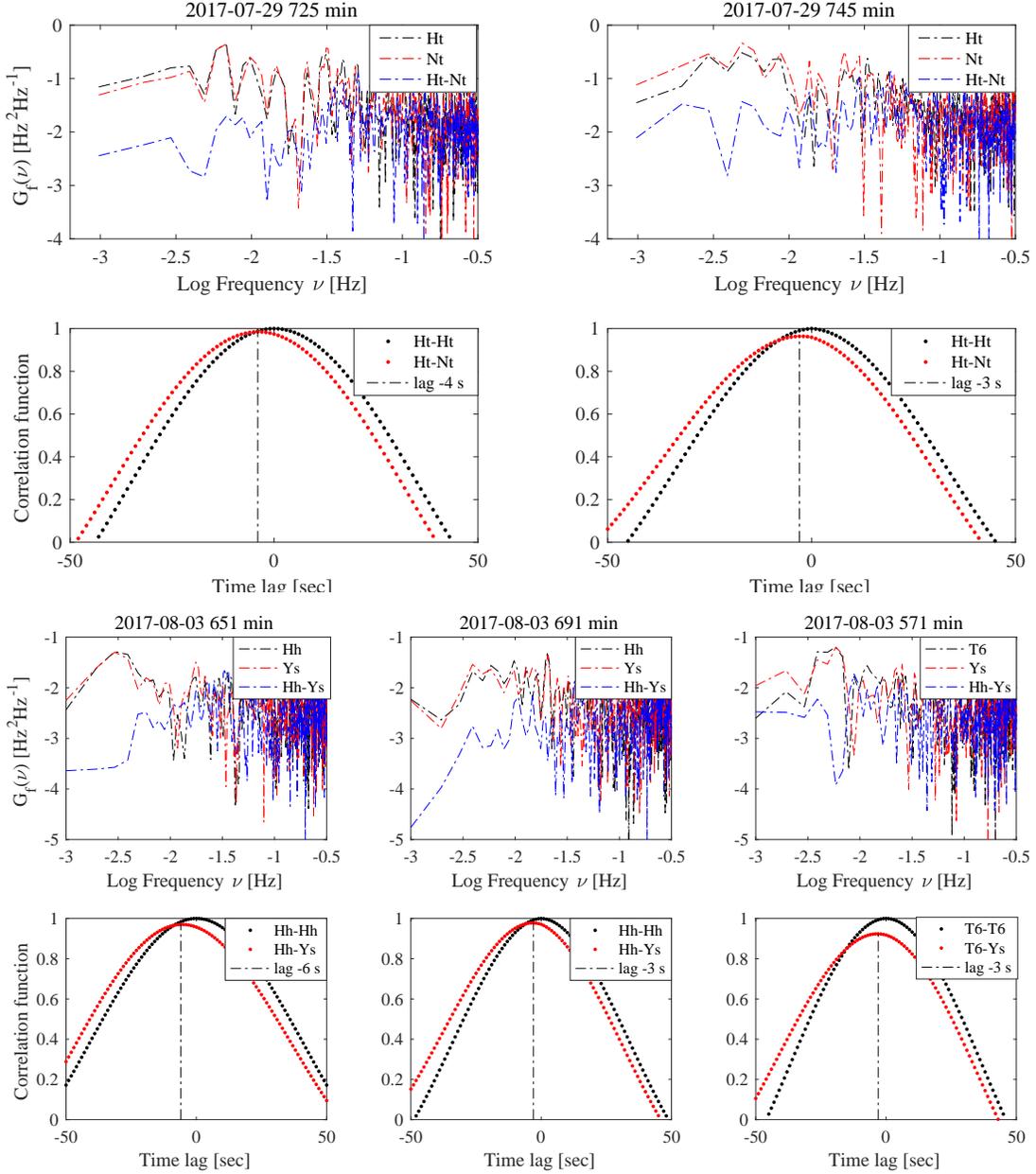

**Figure 9.** The correlation function of fluctuation frequency for two individual scans at the baseline Ht-Nt on 29 July when solar offset was 4.9 R$_s$ and other three individual scans on 3 August when solar offset was 9.9 R$_s$.

**Table 3**. Cross-correlation results of frequency fluctuations on 29 July and 3 August in 2017: Date, the average time at each scan, solar offset, baseline, the correlation coefficients, time lag, the radial projection distance, solar wind velocity, the correlation coefficients, time lag.

| Date | Time | Solar offset | Baseline | C.C [a] | Time lag[a] | $\Delta R$ | $v_c$ | C.C [b] | Time lag[b] |
|------|------|--------------|----------|---------|-------------|------------|-------|---------|-------------|
| | [min] | [$R_s$] | | | [s] | [km] | [km s$^{-1}$] | | [s] |
| 2017-07-29 | 725 | 4.9 | Ht - Nt | 0.98 | -4.0±0.5 | -3800 | 950±120 | 0.96 | -1.0 |
| 2017-07-29 | 745 | 4.9 | Ht - Nt | 0.97 | -3.0±0.5 | -3700 | 1233±220 | 0.94 | -2.0 |
| 2017-08-03 | 651 | 9.9 | Hh - Ys | 0.97 | -5.3±0.5 | -3900 | 735±70 | 0.92 | -1.0 |
| 2017-08-03 | 691 | 9.9 | Hh - Ys | 0.95 | -3.7±0.5 | -4000 | 1080±150 | 0.93 | -2.0 |
| 2017-08-03 | 571 | 9.9 | T6 - Ys | 0.92 | -3.7±0.5 | -3700 | 1000±140 | 0.83 | -2.0 |

[a] The correlation coefficient and time lag from cross-correlation with $\nu_c = 0.01$ Hz.
[b] The correlation coefficient and time lag from cross-correlation with $\nu_c = 0.02$ Hz.

Table 3 also presents the cross-correlation results for a cutoff frequency of $\nu_c = 0.02$ Hz. Both the correlation coefficient and time lag decrease with the increase of $\nu_c$ to 0.02 Hz. The correlation coefficient is approximately 0.6 and the time lag is a bit smaller when $\nu_c = 0.1$ Hz. Janardhan et al. (1998) analysed the effect of the cutoff frequency $\nu_c$ on the cross-correlation of frequency fluctuations of Ulysses when solar offset was 29 $R_s$ in the year 1991. The $\nu_c$ were set to 0.2, 0.125 and 0.025 Hz respectively. The correlation coefficient were 0.62, 0.71 and 0.94, with the corresponding time lag 9.0±0.1 s, 9.3±0.1 s and 10.5±0.1 s,respectively. Therefore, the more powerful filtering steadily improved the correlation coefficient and the time lag. In our case, the smaller solar offsets (4.9 $R_s$ and 9.9 $R_s$) of the radio ray path to MEX bring more noise from the sun, which affect the correlation of turbulence among different LOSs. On the other hand, the application of the X-band (8.4 GHz) tracking system reduces the effect from solar plasma by 16 times compared with S-band (2.2 GHz). Since the noise on the power density spectra of the differential fluctuations above 0.01 Hz is dominant (Fig. 7 and Fig. 9), the results from filtering with $\nu_c = 0.02$ Hz is not used.

### 5.3. *Comparison with other work*

Several spacecraft, including Galileo, Ulysses, Venera-15 and 16 have been used for the estimation of the solar wind velocity $v_c$ by cross-correlation of frequency fluctuations at ground stations spaced at large distance. Values of $v_c$ were between $100 \sim 700$ km s$^{-1}$ using Galileo and Ulysses data when the heliolatitudes of the spacecraft were close to the ecliptic (N.A.Armand et al. 2003; Efimov et al. 2010). The profile of $v_c$ showed that the solar wind acceleration continued at least to solar offset 20 $R_s$. Large $v_c$ were found over the pole regions of the Sun when Ulysses passed through all heliolatitudes from South Pole to equator at distances between $20 \sim 30$ $R_s$ during the conjunction in February $\sim$ March 1995 (Janardhan et al. 1998; Efimov et al. 2010). The $v_c$ exceeded 1000 km s$^{-1}$ in some cases when the heliolatitude was above 70°. Efimov et al. (2010) obtained a relationship between the $v_c$ and the heliolatitude as $v_c(r, \phi) = v_c(r)(1 + 8|\phi|)$, where $\phi$ is the heliolatitude.

Fast $v_c$ were also found from the in-situ observation of $v_c$ by SWOOPS (Solar Wind Observations Over the Poles of the Sun) on Ulysses throughout its two solar polar orbits from 1992 to 2003 (McComas et al. 2000, 2003). The first orbit occurred through the solar cycle declining phase and minimum while the second orbit spanned solar maximum. The solar wind displayed a remarkable simple bimodal structure with persistently fast, uniform solar wind at high heliolatitudes and slower, more variable wind at low latitudes in the first solar polar orbit. The global structure of the 3-D solar wind was completely different during Ulysses second solar orbit, where highly variable flows were observed at all heliolatitudes. The maximum velocity measured by the SWOOPS was about 800 km s$^{-1}$. The correlation of $v_c$ with the heliolatitude was presented by the interplanetary scintillation(IPS) observation of quasars as well (Sokół et al. 2015; Tokumaru et al. 2010). Furthermore, the IPS observations of quasars by STEL(Solar Terrestrial Environment Laboratory) clearly demonstrated that $v_c$ at high latitudes was considerably faster than that at the solar equator during the solar activity declining and minimum phases (Tokumaru et al. 2010).

The $v_c$ measured by the space weather prediction center (SWPC)[4] were 431 km s$^{-1}$ on July 29 and 381 km s$^{-1}$ on August 3 in 2017, significantly smaller than the values obtained from cross-correlation of the frequency fluctuations of MEX on July 29 and August 3. The $v_c$ provided by SWPC were the results from the spacecraft located upwind of the Earth, typically orbiting the L1 Lagrange point. However, the $v_c$ we measured are related to the coronal region with heliolatitudes of the proximate points 70° on July 29 and 40° on August 3 (Fig. 5), respectively. Furthermore, the solar cycle in 2017 was in the declining and minimum phases. According to Tokumaru et al. (2010), $v_c$ at high latitudes was faster than that at the solar equator during the declining and minimum phases. Therefore, the higher heliolatitudes may be the main reason for the outflowing fast solar wind we measured on July 29 and August 3 (Janardhan et al. 1998; Efimov et al. 2010; McComas et al. 2000; Tokumaru et al. 2010). On the other hand, the measured velocities are the instantaneous values estimated from each scan with a duration of 15 minutes. We could not achieve the outer scale of turbulence due to the relatively short length of the time series segments. Efimov et al. (2010) proposed that the speed may exceed the actual solar wind speed due to the instantaneous changes of the inhomogeneities during their motion. Therefore, the longer arc observations are required to study the profile of $v_c$ on a larger scale.

## 6. CONCLUSIONS

This paper has described the data processing of VLBI observations of coronal radio sounding of a spacecraft. In order to obtain the frequency and phase fluctuations, we have adopted two different Doppler compensation methods. One is the PRIDE technique that uses polynomial fits, the other is the local correlation technique that uses *a-prior* ephemeris to measure the fluctuations of the spacecraft carrier signal per each station. We have demonstrated the applications of analysing the frequency fluctuations in the study of plasma turbulence.

We have studied the radial dependence of the frequency fluctuations of ESA's Mars Express in the corona at 4.9 $\sim$ 76.3 R$_s$. The power spectra of the frequency fluctuations show a power-law shape close to Kolmogorov's law at the spectral interval of $\nu_{lo} \sim \nu_{up}$ with a mean spectral index of 0.58. The RMS intensity of the frequency fluctuations $\sigma_f$ follows the power law with the radial solar offset distance.

The frequency fluctuations on June 25, 2015, with a solar offset of 11.2 R$_s$, are stronger than the results at 4.9 R$_s$ on July 29, 2017. A comparison with LASCO images shows that the stronger fluctuations are associated with a bright streamer structure that crossed the LOS during the session.

The electron column density fluctuations, as well as the differential fluctuations along the baseline, are obtained from the phase fluctuations at a solar offset of 4.9 R$_s$ on July 29 and 9.9 R$_s$ on August 3, 2017. The correlation and inhomogeneity of the electron density can be studied by comparing the electron column density fluctuations between different stations. The scale of correlation of the fluctuations is commensurable with the large scale of turbulence with the fluctuation frequency $\nu < 0.01$ Hz. The convective velocities of the irregularities estimated from the cross-correlation of frequency fluctuations are 950 $\sim$ 1233 km s$^{-1}$ on July 29, 755 $\sim$ 1080 km s$^{-1}$ on August 3. The velocities obtained with our method are significantly larger than the measurements made by SWPC. One possible reason is that the solar wind regions related to the measured velocity are different. The fast solar wind velocities correspond to the higher heliolatitude of the coronal region, 70° on July 29 and 40° on August 3, while the velocities providing by SWPC related to the region upwind of Earth, typically orbiting the L1 Lagrange point.

Mars will experience solar conjunction in 2021. Three Mars spacecraft, ESA's Mars Express, ESA's Trace Gas Orbiter and China's Tianwen'1, provide the opportunity to conduct radio science observations of the solar plasma. The VLBI observations and analysis methods during the coronal presented in this paper can be used to study the electron column density fluctuations and the turbulence at multiple spatial points in the inner solar wind by providing multiples lines of sight between the Earth and the spacecraft. The north-south orientation of baseline is particularly sensitive to study solar wind at high latitudes (Rao et al. 2008). A further analysis of Faraday rotation during solar conjunction observation by VLBI network can support the study of the MHD wave processes in the middle-coronal magneto-ionic environment (Wexler et al. 2019b; Jensen et al. 2013).

The authors would like to thank Sheshan 25 m and the European VLBI Network for providing the tracking data used in this paper. We would like to express our sincere gratitude to T.Morley (ESA ESOC) for persistent support. We would also like to thank Prof. Fengchun Shu, Dr. Lei Liu, Zhangjuan and Deng Tao, PhD candidates Zhao Xu, Li

---

Ting for their important support. The author thanks very much the Joint Institute for VLBI ERIC and Prof. Leonid Gurvits for half year's host from 2019-03 to 2019-09. This work began during my stay in JIVE. We would like to thank the National Natural Science Foundation of China (grant numbers 11703070,U1831137,U1931135,11903067,11803070) who funded the project. This study was supported by the National R&D Infrastructure and Facility Development Program of China, 'Fundamental Science Data Sharing Platform' (DKA2017-12-02-09), the Pre-research Project on Civil Aerospace Technologies No. D020303 funded by China National Space Administration.

# APPENDIX

## A. DETERMINATION THE $\nu_{LO}$ FROM THE TWO METHODS FOR COMPENSATING DOPPLER SHIFT

This work uses two different methods to eliminate the Doppler shift. The first method uses an (n-1)-order frequency polynomial fit on the measured Doppler frequency to generate an n - order phase-stopping polynomial fit to stop the Doppler phase (Ma et al. 2017). The second local correlation method uses the precise ephemerides model to shift the Doppler effect (Ma et al. 2020). Here we compare $\delta f$ and $\delta \varphi$ from the two different Doppler compensation methods.

Fig. 10 shows the power density spectra of $\delta f$ from Hh station on 2017-08-03 using the two methods. The spectra are exactly the same when $\nu > 3$ mHz. A small difference appears when $\nu < 3$ mHz. A similar phenomenon appears in the power density spectra of $\delta \varphi$. Fig. 11(a) are the related $\delta \varphi$ at the same scan. Though the $\delta \varphi$ trends are similar, there is still a difference between the two phase time series (Fig. 11(b)). We use higher-order to fit the phase fluctuations difference. However, the residuals of the phase fluctuations difference after fit still have trend terms even with the order-10 (Fig. 11(c)). It indicates there is some complex dynamical motion that could not be fit by a polynomial.

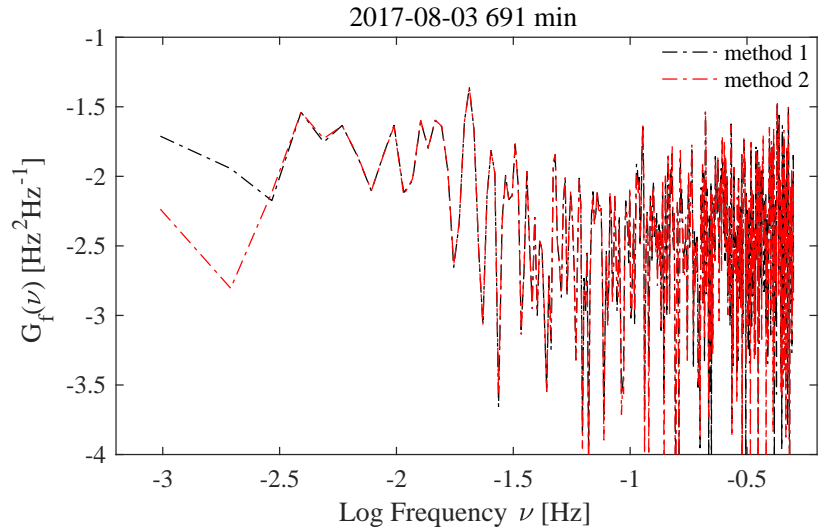

**Figure 10.** The power density spectra of frequency fluctuations at 691 min, 2017-08-03, Hh station. The time is the average of this scan.

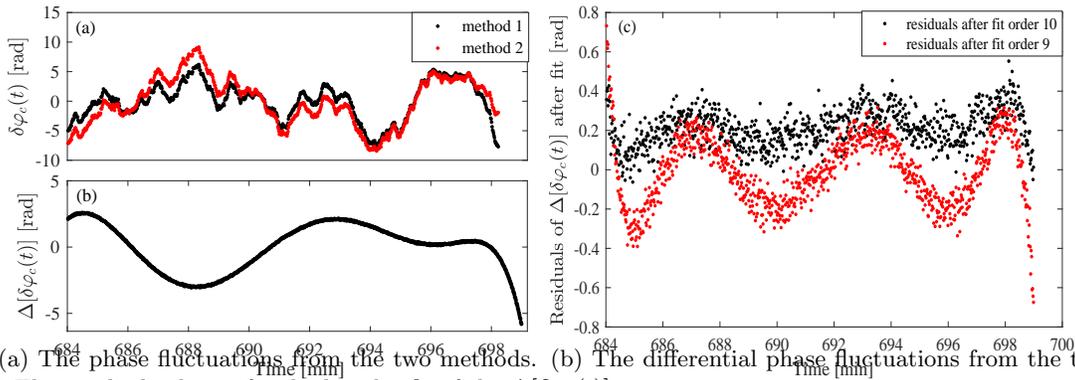

**Figure 11.** (a) The phase fluctuations from the two methods. (b) The differential phase fluctuations from the two methods, $\Delta[\delta \varphi_c(t)]$. (c) The residuals phase after high order fit of the $\Delta[\delta \varphi_c(t)]$.

## B. DETERMINING $\nu_{LO}$ FROM DOPPLER COMPENSATION WITH THE POLYNOMIAL FIT PHASE MODEL

The first method, subtracting the Doppler shift through polynomial fit on the measured topocentric frequency to obtain the fluctuations is very convenient and efficient. The application of the second local correlation method will be limited in the case that no detailed information about the uplink station and transmit frequency. Here we propose an analysis to determine the $\nu_{lo}$ when only the first method available.

Fig. 12(a) is an example of one scan of the Doppler frequency data on 2017-07-07 for MEX at Sh station. The length of the time series is 19 mins. We use the different 3-order, 4-order, 5-order, 6-order to remove the trends and obtain the frequency fluctuations (Fig. 12(b)), then analysis the power density spectra of the fluctuations as Fig. 12(c). The spectra are different when $\nu < 3$ mHz, thus $\nu_{lo} \approx 3$ mHz, similar to the result determined from Appendix A. Therefore, though the low-frequency plasma fluctuations is sensitive to the polynomial fit, we can ensure the validity when $\nu > \nu_{lo}$, $\nu_{lo} \approx 3$ mHz.

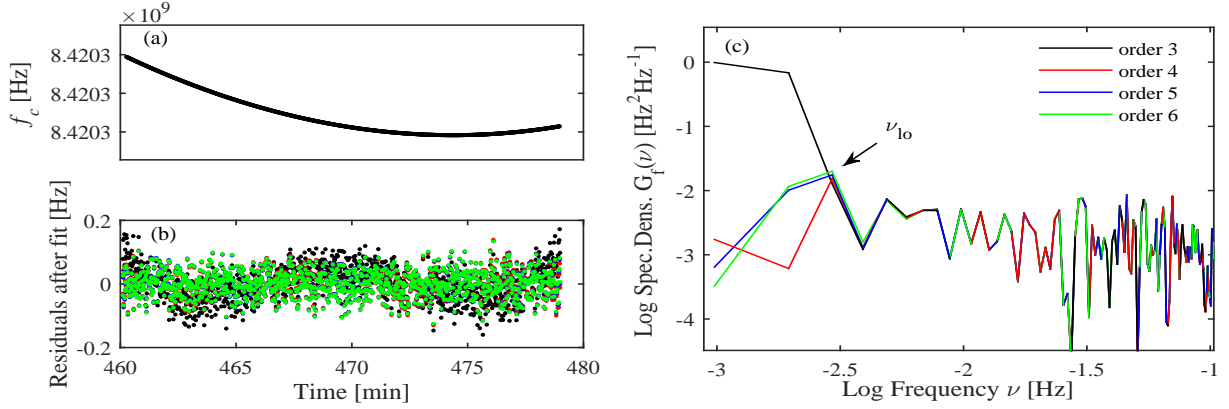

**Figure 12.** (a) The measured frequency of the main carrier. (b) The residuals after different order fit. (c) The spectra of the residuals frequency. The spectra are different at $\nu < \nu_{lo}$, $\nu_{lo} \approx 3$ mHz.